# Gigagauss magnetic field generation by bladed microtube implosion

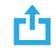

View Online   Export Citation   CrossMark

D. Pan and M. Murakami[a]

**AFFILIATIONS**
Institute of Laser Engineering, Osaka University, Suita, Osaka 565-0871, Japan

[a]Author to whom correspondence should be addressed: murakami.masakatsu.ile@osaka-u.ac.jp

**ABSTRACT**

We demonstrate the generation of ultrahigh magnetic fields—in the order of gigagauss—using a bladed microtube target whose inner surface is periodically slanted in a sawtooth-like pattern. When irradiated by ultra-intense, ultrashort laser pulses, hot electrons with MeV energies are produced at the outer surface and swiftly transported to the inner surface, initiating a rapid implosion of plasma toward the central axis. The unique blade-induced asymmetry gives rise to vortex-shaped flows of ions and electrons near the center, forming strong azimuthal loop currents that generate ultrahigh magnetic fields at the center. Two-dimensional particle-in-cell simulations, supported by a simple analytical model, elucidate the underlying physics and reveal key scaling laws governing the field strength and spatial confinement.

## I. INTRODUCTION

Ultrahigh magnetic fields play a crucial role in high-energy-density (HED) physics and laboratory astrophysics, enabling the study of extreme astrophysical phenomena,[1,2] enhancing plasma confinement in fusion applications,[3,4] and facilitating high-energy particle acceleration and radiation generation.[5–7]

Thanks to advances in ultra-intense femtosecond laser technology[8,9] and precision micro-fabrication,[10,11] various laser-driven magnetic field generation schemes have been developed, such as curved targets,[12,13] capacitor-coil targets,[14,15] and microtube implosion schemes.[16–18] These approaches have enabled magnetic field strengths ranging from tens of tesla to several hundred kilotesla.

Among them, the microtube implosion (MTI) scheme, illustrated in Fig. 1, has demonstrated the ability to produce megatesla-level magnetic fields.[16] In this scheme, a hollow cylindrical target with inner radius $R_0 \sim 1\text{--}10\,\mu\text{m}$ is irradiated by ultra-intense femtosecond laser pulses ($I_L \sim 10^{20}\text{--}10^{22}\,\text{W/cm}^2$), producing hot electrons of MeV energies. These hot electrons form a sheath field along the inner wall, which accelerates ions radially inward (implosion). A seed magnetic field deflects the ions and electrons in opposite azimuthal directions via the Lorentz force, inducing loop currents ($J_{i\phi}$ and $J_{e\phi}$) in the same direction, ultimately generating a strong axial magnetic field $B_c$. While effective, this scheme requires a kT-level seed field, introducing system complexity and limiting compact implementation.

To overcome these limitations, we propose a new concept: the bladed microtube implosion (BMI), illustrated in Fig. 2. The BMI scheme employs a hollow cylindrical target with a periodically slanted inner surface resembling sawtooth-shaped blades. This design breaks the cylindrical symmetry of the imploding plasma, causing azimuthally asymmetric ion acceleration and generating a spontaneous loop current around the center.[19] This self-generated current gives rise to a gigagauss-level axial magnetic field, even in the absence of an externally applied seed field.

The acceleration mechanism underlying BMI is conceptually akin to target normal sheath acceleration (TNSA), where laser-driven hot electrons rapidly escape into the central vacuum, creating sheath fields that accelerate ions normal to the inner surface. In BMI, the blade geometry redirects these ions slightly off-axis, inducing a net loop current that seeds the magnetic field. This geometry-driven anisotropy enables spontaneous field generation through collective plasma dynamics.

Unlike conventional flux compression schemes[20–22] that rely on compressing an initially applied seed magnetic field, the present bladed microtube implosion (BMI) concept fundamentally differs in its mechanism. Here, no external magnetic field is required at the outset. Instead, a spontaneous loop current is self-consistently generated near the center due to the asymmetric ion acceleration induced by the blade geometry. This loop current acts as an intrinsic seed, which is then amplified through collective plasma motion during the implosion process. The resulting magnetic field emerges from zero and grows rapidly to gigagauss levels, marking a distinct paradigm shift from seed-field compression to geometry-driven magnetic field creation.

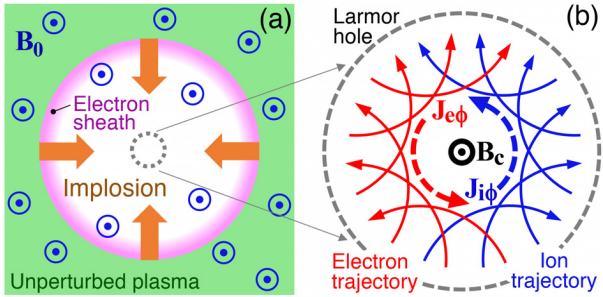

FIG. 1. (a) Top view of inner plasma dynamics during microtube implosion. Isothermal expansion of the inner-wall plasma into vacuum is driven by laser-produced hot electrons. (b) An ultrahigh magnetic field $B_c$ forms at the center, induced by collective currents of ions and electrons deflected in opposite directions by the seed field $B_0$. Reproduced from Murakami et al., Sci. Rep. **10**, 16653 (2020); licensed under a Creative Commons Attribution (CC BY) license.

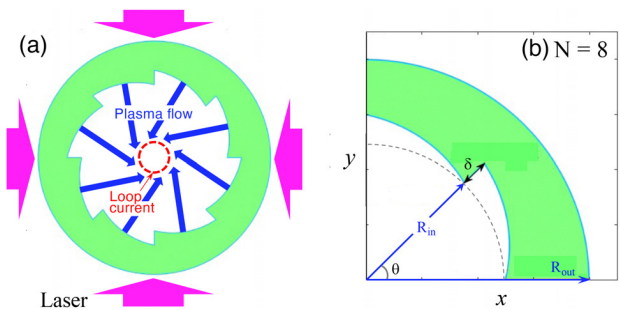

FIG. 2. (a) Core mechanism of BMI: blade heating by hot electrons drives azimuthal ion motion, inducing a loop current and ultrahigh magnetic field. (b) Example with eight blades.

The remainder of this paper is organized as follows: Sec. II introduces the bladed microtube concept and the mechanism of spontaneous magnetic field generation, with emphasis on key simulation results and contrasts with conventional compression schemes. Section III presents an analytical model to derive scaling laws for the magnetic field and core size, including polarity transition. Section IV discusses blade number optimization, 3D stability, and applications. Section V concludes the paper.

## II. PIC SIMULATIONS OF BLADED MICROTUBE IMPLOSION
### A. Setup and laser configuration

The blade structure is intended to provide non-uniform initial conditions for the ion and electron flow under laser irradiation by geometrically breaking the symmetry, thus facilitating the spontaneous generation and amplification of the azimuthal electric current and magnetic field. Below, we perform 2D PIC simulations of BMI using the fully relativistic open-source code EPOCH.[23] The simulation box placed on the x–y plane has a size of $22\,\mu m \times 22\,\mu m$ at a rate of 100 cells/$\mu$m or equivalently 10 nm/cell. The system is assumed to be uniform along the z axis. Carbon with an initial density of $n_{i0} = 3 \times 10^{22}$ cm$^{-3}$ is employed as the target material in this study, which is assumed to be fully ionized at the ionization state $Z = 6$, and the corresponding initial electron density is therefore $n_{e0} = 1.8 \times 10^{23}$ cm$^{-3}$. Each square cell for the target material is filled with 100 pseudo ions and 200 pseudo electrons.

In our simulations, both ions and electrons are treated as fully kinetic particles using the standard particle-in-cell (PIC) method implemented in the EPOCH code. In the simulations, a realistic proton-to-electron mass ratio $m_p/m_e = 1836$ is employed (the mass of a carbon ion is $m_i = 12m_p$). Binary collisions and ionization processes are neglected in this study, since the plasma is considered collisionless over the short laser interaction timescale ($\sim 100$ fs). These approximations are justified by the relativistic energies of the electrons and the extremely high temperature of the system, where collective electromagnetic effects dominate the particle dynamics.

The target structure of BMI is characterized by four free parameters, i.e., the number of blades $N$, the innermost radius $R_{in}$, the outer radius $R_{out}$, and the gap on the blade edge $\delta$. In Fig. 2(b), $N = 8$, $R_{in} = 5\,\mu m$, $R_{out} = 8\,\mu m$, and $\delta = 1\,\mu m$ are employed as an example set of the parameters. The elementary shape of the bladed inner surface is given by a sinusoidal curve in this study, i.e., $r(\theta) = R_{in} + \delta \sin(N\theta/4)$ for $0 \leq \theta \leq 2\pi/N$, which is periodically assigned along the inner circle ($0 \leq \theta \leq 2\pi$) to form N-blades.

Four linear-polarized laser pulses with a laser wavelength of $\lambda_L = 0.8\,\mu m$ and a peak intensity of $I_L \sim 10^{21}$ W/cm$^2$ propagate toward the BMI target located at the center along the $\pm x$- and $\pm y$-axes. The polarization of the incident laser pulses is specified such that the electric field and the magnetic field components oscillate on the x–y plane and along the z axis, respectively. The temporal profiles of the incident laser pulses have Gaussian distribution with the pulse duration $\tau_L = 100$ fs (FWHM: full width at half maximum), while the transverse profiles of the pulses are assumed to be plane waves.

### B. Implosion dynamics

Figures 3(a)–3(d) and 3(e)–3(h) depict the temporal evolution of ion and electron density distributions, respectively, at four representative time points: $t = 125$, 240, 300, and 600 fs. These snapshots reveal the dynamic progression of the system throughout the implosion process. Upon irradiation by intense laser pulses, hot electrons generated at the outer surface rapidly traverse to the inner surface of the microtube and penetrate into the central vacuum region of the hollow target. This inward electron flux forms a sheath that induces a quasistatic electric field near the inner wall, which in turn accelerates wall ions toward the central axis [Figs. 3(a), 3(b), 3(e), and 3(f)]. Notably, as seen in Fig. 3(b), the resulting ion flow exhibits a symmetric off-center pattern, consistent with expectations based on the target geometry. As the imploding plasma front, composed of both ions and electrons, passes near the center of the target, a characteristic ring structure—referred to as a "Larmor hole"—emerges around the central region [Figs. 3(c) and 3(g)]. The Larmor hole typically measures 1–2 $\mu m$ in diameter and represents the envelope of the gyro-orbits of charged particles. In a conventional microtube target without any inner-wall blade structure, the Larmor hole forms due to the deflection of particles by seed magnetic fields on the order of kilotesla.[16] In contrast, in the BMI target, it arises from the geometrical asymmetry introduced

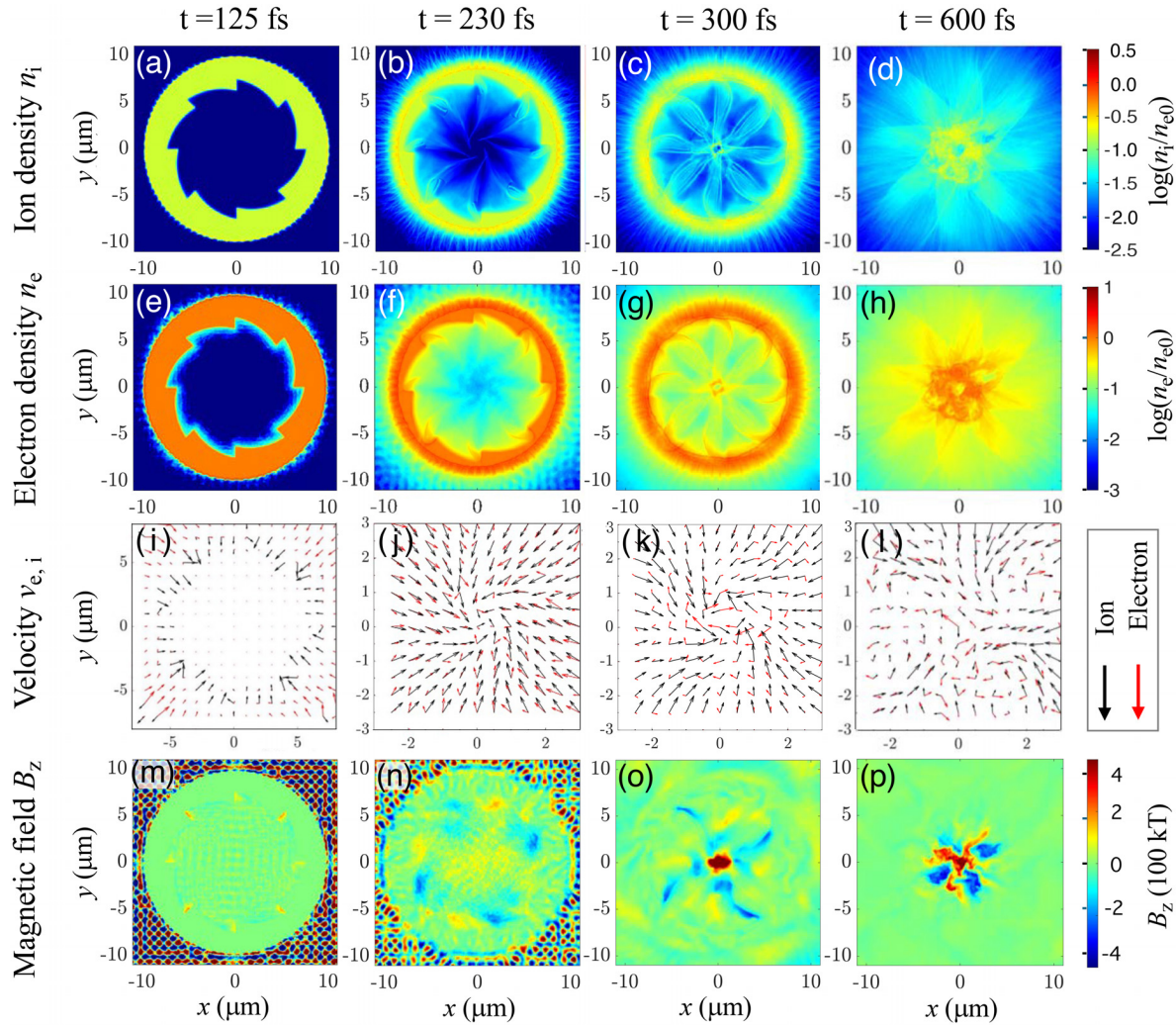

**FIG. 3.** Two-dimensional distributions of key physical quantities at four different times, under a laser pulse duration of $\tau_L = 100$ fs and peak time $t_p = 150$ fs. (a)–(d) Ion density $n_i$. (e)–(h) Electron density $n_e$. (i)–(l): Velocities of ions (black arrows) and electrons (red arrows), showing anticlockwise ion motion and clockwise electron motion in the central region, resulting in the same anticlockwise current generation. This converging vortex pattern forms the core of the stagnation structure. (m)–(p) Axial magnetic field $B_z$.

by the blades, which induce a pronounced vortex-like motion of ions and electrons during the implosion.

Figures 3(i)–3(l) and 3(m)–3(p) show the temporal evolution of the ion and electron velocity fields and the magnetic field distributions, respectively. At the peak time of the magnetic field, as illustrated in Fig. 3(k), vortex motions of the ions and electrons are clearly observed—anticlockwise for ions and clockwise for electrons. This converging vortex configuration, consisting of clockwise electron motion and counterclockwise ion motion, forms a quasi-stationary current loop centered near the axis. This self-organized flow pattern underlies the magnetized stagnation structure clearly observed in Fig. 3(o). These counter-rotating flows act cooperatively to generate a strong azimuthal loop current on the order of peta-amperes per square centimeter in the anticlockwise direction. This intense current gives rise to a localized magnetic field of approximately 500 kT at the center, confined within an ellipsoidal region with major and minor axes of $2\,\mu$m and $1\,\mu$m, respectively, as seen in Fig. 3(o). It is worth noting that the formation of such vortex structures and the associated magnetic field arises through a self-consistent positive feedback mechanism. Specifically, the initial loop current amplifies the central magnetic field, which in turn constrains the motion of charged particles more tightly via the Lorentz force—thereby reinforcing and further intensifying the loop current itself.

### C. Magnetic field formation

Figure 4 shows the energy spectra of ions and electrons at two different times before ($t = 200$ fs) and after ($t = 300$ fs) the collapse of converging plasma at the center. The electron energy spectra at both times are found to be well fitted by the Maxwell distribution with a temperature of $T_e \simeq 13$ MeV (compare the fitted curve). This electron temperature can be well explained by the ponderomotive scaling,[24] i.e., $T_e(\text{MeV}) \approx 14\sqrt{I_{L21}\lambda_{L\mu m}^2}$, where $I_{L21}$ and $\lambda_{L\mu m}$ are the laser intensity and the laser wavelength in units of $10^{21}$ W/cm² and $\mu$m, respectively. The scaling reads $T_e \sim 11$ MeV by applying $I_L = 10^{21}$ W/cm² and $\lambda_L = 0.8\mu$m. In addition, the high-energy tail of the ion distribution is also found to have a similar temperature, $T_i \simeq 13$ MeV. While the ion temperature and the electron temperature are kept close to each other in the implosion phase ($t = 200$ fs), i.e., $T_e \approx T_i$, the kinetic energy of an imploding ion increases in time due to adiabatic expansion. The implosion velocity $v_{imp}$ is then roughly estimated to reach the order of the sound speed, i.e., $v_{imp} \sim (1-2)c_s \sim (2.5$–$5.0) \times 10^9$ cm/s, where the sound speed $c_s = \sqrt{ZT_e/m_i}$ is calculated with the aid of $T_e = 13$ MeV. The estimates well explain the obtained simulation result, $v_{imp} \sim 4 \times 10^9$ cm/s, and the corresponding kinetic energy of carbon ions $\mathcal{E}_{imp} \sim 100$ MeV.

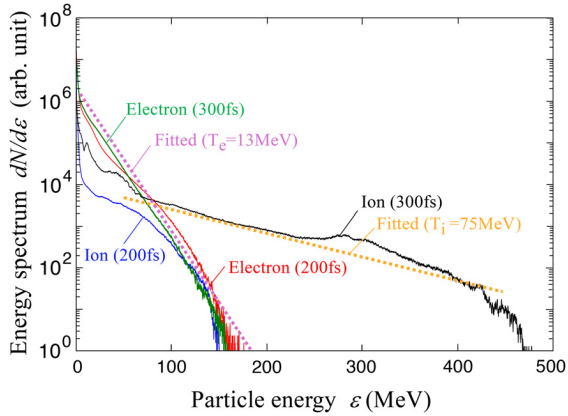

FIG. 4. Energy spectra for ions and electrons, corresponding to the two different times—slightly before ($t = 200$ fs) and slightly after ($t = 300$ fs) the rising of the strong magnetic field at the center. Fig 3 shows the corresponding 2D maps of key physical quantities at the two respective times.

Meanwhile, the ion energy spectrum at $t = 300$ fs in Fig. 4, corresponding to $T_i \approx 75$ MeV (compare the fitted curve) for the energy range $100\,\text{MeV} \lesssim \varepsilon \lesssim 400\,\text{MeV}$, makes a striking contrast to the ion spectrum at $t = 200$ fs. This strong heating of ions and the generation of a strong magnetic field at the center coherently occur. In other words, when the imploding ions are passing by the closest points to the center, the local density quickly increases to reach the same order as the solid density, and at the same time strong currents induced by both the ions and electrons generate the ultrahigh magnetic fields. The magnetic fields thus generated then trap the ions, which convert a substantial amount of their own imploding kinetic energy into thermal energy via ion–ion collisions. As a result, the ions remain tightly trapped by the self-generated magnetic field in the order of sub-megatesla and are confined in this region for a duration exceeding a few picoseconds, which is substantially longer than the pulse duration of the applied laser ($\sim 100$ fs).

The generation and saturation of the axial magnetic field can be interpreted as a feedback loop: ion and electron flow induced by the blades drive azimuthal currents, which generate axial magnetic fields, compressing the plasma further. This positive feedback continues until the system reaches a saturated state, as observed in Fig. 3(o). The self-organization of the loop current relies critically on the geometrical asymmetry introduced by the blades.

As the self-generated axial magnetic field increases, the Larmor radii of both electrons and ions shrink, confining their transverse motion more tightly around the axis. This magnetic confinement effectively compresses the plasma into a narrower region, thereby increasing the local current density and enhancing the magnetic field in return.

The saturation of the axial magnetic field results from several interlinked mechanisms: the finite capacity of hot electrons to sustain azimuthal currents, magnetic mirror effects that hinder electron circulation, and opposing electrostatic fields arising from ion return flows. Additionally, increased confinement leads to energy concentration, which may trigger wave excitation or turbulent dissipation, disrupting the current loop structure. These effects collectively halt the further amplification of the magnetic field.

Unlike conventional Z-pinches, where external currents generate azimuthal magnetic fields that pinch the plasma radially, the present mechanism is entirely self-generated. The axial magnetic field arises from asymmetric, laser-driven particle flows, and the resulting compression is not driven by explicit magnetic pressure but emerges from kinematic confinement due to reduced Larmor motion. This represents a fundamentally different type of self-induced magnetic compression.

### D. Blade number optimization

It is of significant interest to investigate the effect of the number of blades, $N$, on the generation of magnetic fields. Figure 5 shows temporal evolutions of maximum magnetic field along the z axis $B_c$ in the central area for different numbers of $N$. Note that the temporal evolution for $N = 8$ in Fig. 5 corresponds to the case given in Fig. 3. As can be seen in Fig. 5, the enhancement and retention of the magnetic field over time is most significant with $N = 8$, where not only the peak value is high but also the strength remains relatively stable during the decay phase. Here, it should be noted that the reason for the highest performance with $N=8$ seems to be strongly related to the fact that the target is orthogonally illuminated by four laser beams. Meanwhile, too low or too high a number of blades is found to be detrimental to the self-consistent amplification of the magnetic field during the implosion phase.

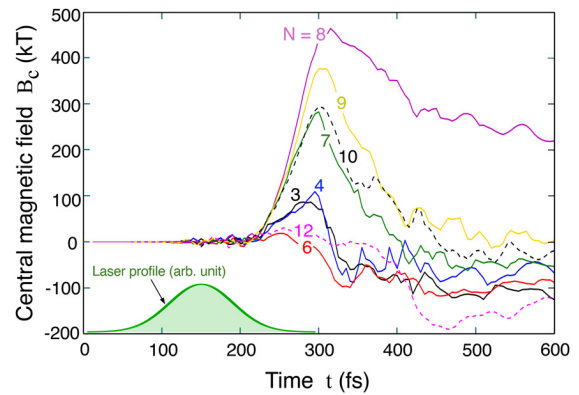

FIG. 5. Temporal evolution of the central magnetic field $B_c$ at the center for various blade numbers $N$. The case with $N = 8$ yields the strongest and most sustained magnetic field, highlighting the importance of optimal target geometry for field amplification.

These simulation findings point to a coherent underlying mechanism. In what follows, we develop a simple analytic framework to capture the essential scaling behavior of the generated magnetic field.

## III. ANALYTICAL INTERPRETATION AND SCALING LAW
### A. Larmor hole and magnetic field

To interpret the simulation results from a more quantitative perspective, we now present a simple analytical model based on scaling arguments. In particular, we aim to explain the magnetic field generation mechanism observed in Fig. 3 and its parameter dependence.

The typical size of a Larmor hole in the BMI system, which is formed under a self-regulated magnetic field $B_c$, can be roughly estimated as follows. Suppose that a single ion with a mass $m_i$, an ionization state $Z$, and a velocity $v_i$ is moving under a uniform magnetic field $B_c$. The Larmor hole radius $r_H$ is expected to be in the same order as the corresponding Larmor radius, i.e., $r_H \sim r_L = m_i v_i / ZeB_c$, where $e$ is the elementary charge. Meanwhile, when a uniform ion flux of the same kind with a number density $n_i$ is rotating on a circle with radius $r_L$, the induced magnetic field in the center is estimated, based on Ampère's law, to be $B_c \sim 4\pi j_i r_L / c^2$ with the ion current density $j_i = Zen_i v_i$, where $c$ is the speed of light. From these relations, one obtains the Larmor hole radius,

$$r_H \sim \sqrt{\frac{m_i c^2}{4\pi Z^2 n_i e^2}}, \quad (1)$$

and the magnetic field at the center,

$$B_c \sim \frac{\sqrt{4\pi n_i T_e}}{c}, \quad (2)$$

where we assumed $m_i v_i^2 \sim T_e$.

Using typical values obtained from our simulations—namely, ion density $n_i \sim (1-2) \times 10^{23}$ cm$^{-3}$ and hot electron temperature $T_e \sim 13$ MeV as inferred from Fig. 4—we estimate $B_c \sim 360$–510 kT. This estimate agrees remarkably well with the simulated peak value of $\sim 500$ kT observed in Fig. 3(o) and Fig. 5 around $t = 300$ fs.

Equation (2) also provides useful insight into optimization. Since $T_e$ depends on laser intensity via the ponderomotive scaling (i.e., $T_e \propto \sqrt{I_L \lambda_L^2}$), and $n_i$ reflects the degree of plasma compression, $B_c$ can be enhanced by (i) increasing laser intensity, (ii) choosing a wavelength to optimize electron heating, and (iii) tailoring the target geometry to increase central ion density. These considerations align with the findings of Fig. 5, which shows that the blade number $N = 8$ yields the strongest and most stable magnetic field.

Although no explicit dimensionless parameter is introduced here as in our previous MTI work,[16] Eq. (2) serves as a compact yet effective scaling law to guide future optimization strategies for blade-based targets in the absence of seed magnetic fields.

### B. Laser intensity scaling

To investigate the laser intensity scaling on the magnetic field generated around the center, $B_c$, as one of the most crucial aspects of BMI study, we performed a series of simulations, under the variance, $2.5 \times 10^{20} \lesssim I_L$ (W/cm$^2$) $\lesssim 4 \times 10^{21}$. As in Fig. 3, the target parameters, $R_{in} = 5 \mu$m, $R_{out} = 8 \mu$m, $\delta = 1 \mu$m, and $N = 8$, are employed, while we use four Gaussian beams, which have $\tau_L = 100$ fs and are spatially planar waves just for simplicity.

Figure 6 shows the simulation results (red circles) compared with the model prediction (blue dashed line), which is obtained by rewriting Eq. (2) with the aid of $n_i \sim 2n_0 \approx 10^{23}$ cm$^{-3}$ and the ponderomotive scaling, $T_e(\text{MeV}) \approx 14 \sqrt{I_{L21} \lambda_{L\mu m}^2}$, in the form

$$B_c(\text{kT}) \sim 530 \left(I_{L21} \lambda_{L\mu m}^2\right)^{1/4}. \quad (3)$$

While Eq. (3) predicts a scaling of $B_c \propto I_L^{1/4}$ that agrees well with the simulation results for $I_L \gtrsim 10^{21}$ W/cm$^2$, noticeable deviations emerge at lower intensities. This discrepancy arises because Eq. (3) assumes a regime in which the laser pulse generates a sufficiently hot and dense electron population, capable of sustaining a strong sheath field of order $E \sim T_e/\lambda_D$. This electrostatic sheath drives inward ion motion and facilitates the formation of a circulating azimuthal current within the cavity. The resulting closed-loop electron current induces a strong axial magnetic field through Ampère's law. However, in the lower intensity regime ($I_L \lesssim 10^{21}$ W/cm$^2$), the electron temperature $T_e$ is too low to establish a strong sheath, thereby weakening both the azimuthal current circulation and the associated magnetic field generation.

### C. Polarity transition and field reversal

There is another feature of high interest observed in Fig. 5, i.e., the transition of magnetic polarity from $B_c > 0$ to $B_c < 0$. The polarity transition occurs as a result of a complex interplay between the electrons and the electromagnetic fields surrounding them around the center.[17,19] The observed polarity reversal does not occur during the laser interaction but emerges in the post-irradiation stage. Since the laser intensity is fixed in Fig. 5, the triggering factor for reversal is not the laser strength itself but rather the resulting average energy and phase-space spread of the hot electrons.

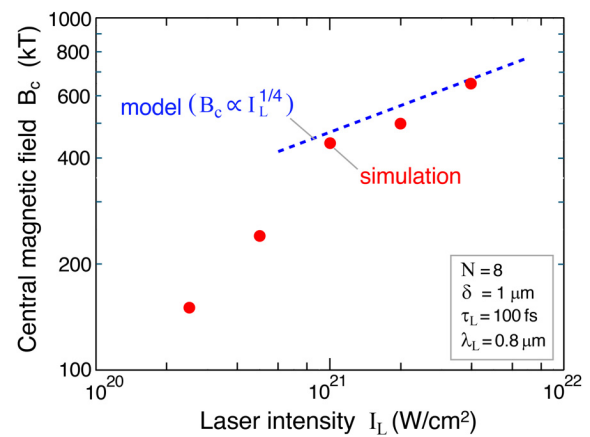

**FIG. 6.** Scaling of the central magnetic field $B_c$ with laser intensity $I_L$. The model prediction [blue dashed line, Eq. (3)] turns out to reproduce well the simulation results (red circles) for $I_L \gtrsim 10^{21}$ W/cm$^2$.

When the average electron energy is moderate, the spread in azimuthal angles becomes sufficiently large that both clockwise and counterclockwise current loops can form. The final polarity of the axial magnetic field reflects which direction becomes dominant over time. This scenario can be interpreted as a dynamic phase-space competition between multiple current channels seeded by hot electrons with diverse injection angles and momenta. Below, we briefly demonstrate, in terms of a simplified numerical model, that opposite directions of electron currents, either clockwise or anticlockwise around the center, can occur by slight changes of a kinetic condition of a test electron and an electromagnetic field acting on it.

Suppose that a charged test particle is put in a uniform electric field $E_0$ and uniform magnetic field $B_0$. As is well known, the resultant drift velocity is given by $v_d = E_0 \times B_0/B_0^2$ with $B_0 = |B_0|$, which is a translational movement, and its direction does not depend whether the charge is positive or negative. In case of a polar electric field instead of a uniform electric field, however, the resultant movement of a test particle becomes polar precession, the direction of which depends on specific conditions as follows.

Without losing the physical essence, we consider a nonrelativistic electron dynamics moving on the $x$–$y$ plane with its temporal position $r = (x(t), y(t), 0)$ under a uniform background magnetic field along the $z$ axis, $B_0 = (0, 0, B_0)$, and a spherically symmetric Coulomb electric field induced by a uniform ion sphere located at the center, $E = E(r)i_r$, where $i_r = r/|r|$. The equation of motion of a test electron is described by $dp_e/dt = -e(E + v_e \times B)$, where $p_e$ and $v_e = (v_x, v_y, 0) = (\dot{x}, \dot{y}, 0)$ are the momentum and velocity vectors of the test particle, respectively. The equation of motion on the $x$–$y$ plane is further simplified in the form

$$\frac{d\tilde{v}_x}{dt} = -\tilde{Q}_0 \tilde{E}_x - \tilde{B}_0 \tilde{v}_y, \quad (4)$$

$$\frac{d\tilde{v}_y}{dt} = -\tilde{Q}_0 \tilde{E}_y + \tilde{B}_0 \tilde{v}_x, \quad (5)$$

where the tilde stands for appropriate normalization to make all the variables, $\tilde{v}_x$, $\tilde{v}_y$, $\tilde{E}_x$, and $\tilde{E}_y$, dimensionless as well as the constants $\tilde{Q}_0$ and $\tilde{B}_0$. The uniform ion sphere with a unit radius is to mimic the imploded ions at the center, which is defined by

$$(\tilde{E}_x, \tilde{E}_y) = \begin{cases} (\tilde{x}, \tilde{y}), & 0 \leq \tilde{r} \leq 1, \\ (\tilde{x}, \tilde{y})\tilde{r}^{-3/2}, & \tilde{r} \geq 1, \end{cases}$$

where $\tilde{r} = \sqrt{\tilde{x}^2 + \tilde{y}^2}$ is the dimensionless distance between the test electron and the center.

Figures 7(a) and 7(b) show that the average electron momentum is clockwise and anticlockwise, respectively, under the specific values for the constants, $\tilde{B}_0$, $\tilde{Q}_0$, and $\tilde{v}_{y0} = \tilde{v}_y(0)$, given in the respective figures, where the other initial conditions, $\tilde{x}(0) = -5$, $\tilde{y}(0) = 1$, and $\tilde{v}_x(0) = 1$, are fixed just as an example. The blue circle stands for the ion sphere. In Fig. 7(a), the resultant magnetic field induced by the test particle, $B_{ind}$, is in the same direction of the uniform background $B_0$, i.e., $B_0 \cdot B_{ind} > 0$. On the contrary, in Fig. 7(b), $B_0 \cdot B_{ind} < 0$. Once the balance of hot electrons moving in the two rotational directions is lost, the polarity transition happens in a cascade-like manner.

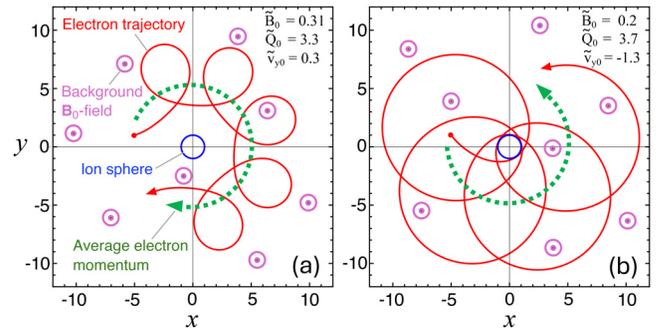

**FIG. 7.** Demonstration of clockwise and anticlockwise electron motion using a simplified numerical model. In (a), the induced magnetic field $B_{ind}$ aligns with the background field $B_0$; in (b), it reverses. This illustrates that the polarity transition can be triggered by the interplay between hot electrons and the electromagnetic fields near the center.

## IV. DISCUSSION

### A. Blade number and geometric optimization

The reason why $N = 8$ yields the strongest magnetic field can be attributed to a geometric resonance between the blade arrangement and the four-beam laser configuration. When $N = 8$, each laser beam illuminates a pair of blades placed symmetrically across the axis (e.g., at $\pm 45°$, $\pm 135°$, etc.), ensuring balanced and constructive electron heating. This promotes synchronized sheath formation and coherent azimuthal current generation.

In contrast, when $N = 4$, the number of blades is too small to break the cylindrical symmetry sufficiently, resulting in a weak loop current. When $N = 12$ or more, each beam interacts with too many small sectors, fragmenting the heating and degrading the coherence of the imploding ion flows. These effects collectively reduce the azimuthal current density and the resultant magnetic field strength.

Thus, $N = 8$ represents an optimal balance where geometric asymmetry is strong enough to drive vortex-like motion, while still maintaining phase coherence of current generation across the target. This explains the simulation trend seen in Fig. 5.

### B. Conceptual connection to previous implosion studies

Our proposed bladed microtube implosion (BMI) scheme exhibits notable parallels and contrasts with several implosion configurations previously studied in the context of inertial confinement fusion (ICF) and Z-pinch systems. For instance, Peterson et al.[25] and Thomas et al.[26] performed 2D R-Z radiation-hydrodynamic simulations of ICF implosions exhibiting oval stagnation regions characterized by quadrupolar vortex flows. Although these configurations were driven by radiation or thermal pressure rather than magnetic forces, they developed self-organized vortex structures and generated strong magnetic fields in the gigagauss range.

Similarly, in magnetically driven Z-pinch systems, dipole-like vortex stagnation patterns have been observed both numerically and experimentally,[27–29] where the magnetic pressure directly drives the implosion. These flows also form large-scale self-organized structures, especially when the applied magnetic field possesses helicity.

In contrast, our BMI approach generates the magnetic field spontaneously during the implosion through geometry-induced azimuthal currents, without relying on an externally imposed magnetic field.

Importantly, while conventional Z-pinch and ICF systems often require external symmetry control or magnetic twist to stabilize the plasma and form coherent structures, the BMI scheme achieves this through intrinsic geometrical asymmetry and feedback. This marks a conceptual departure from field compression-based magnetic implosions and positions BMI as a compact alternative for generating ultra-high magnetic fields and studying magnetized stagnation.

### C. Outlook for applications and future work

Building upon the conceptual advances highlighted in Sec. IV B, the present BMI scheme offers fertile ground for future explorations in both basic and applied plasma physics. Its ability to generate ultrahigh magnetic fields via self-organized current loops—without relying on externally imposed seed fields—opens the door to compact, scalable platforms for studying magnetized high-energy-density matter under controlled laboratory conditions.

To assess the full three-dimensional (3D) stability of the BMI-generated stagnation and magnetic field structure, future investigations involving helical blade designs and full 3D simulations are warranted. As known from fast ignitor and Z-pinch studies, the absence of twist or helicity often leads to plasma instability in the axial direction, causing loss of confinement ("squirting out").[30] Introducing a controlled helical asymmetry into the blade geometry may provide the required topological constraint to stabilize the plasma and sustain the self-organized loop current in 3D. Similar stabilization mechanisms have been observed in helical jet formation and magnetized plasma gun experiments.[31,32] This motivates a natural extension of the current work toward 3D PIC simulations using helically twisted targets. A similar concept of symmetry-breaking-driven magnetic field generation has been discussed in other plasma contexts.[33]

## V. CONCLUSION

We have proposed and validated a novel target design—the bladed microtube (BMI) scheme—for the generation of gigagauss-level magnetic fields using laser–plasma interaction. The introduction of periodic blade structures induces a geometric asymmetry that breaks the symmetry of electron and ion flows, resulting in the formation of a strong loop current and the subsequent generation of an intense axial magnetic field.

Particle-in-cell (PIC) simulations have demonstrated that this geometry-driven mechanism can self-consistently amplify magnetic fields up to 450 kT without requiring any external seed field. An analytic model was developed to interpret the underlying dynamics and predict scaling laws, revealing a feedback loop between magnetic confinement and current enhancement. The analysis also suggests optimization strategies for maximizing field strength through target geometry.

This BMI concept opens a new path to compact, laser-driven, strongly magnetized plasmas, with potential applications ranging from inertial confinement fusion and laboratory astrophysics to high-field materials science. The present study thus lays a robust theoretical and computational foundation for future experimental implementations.

## ACKNOWLEDGMENTS

This work was supported by the Japan Society for the Promotion of Science (JSPS) and the Kansai Electric Power Company, Incorporated (KEPCO). All the computations were performed by using the supercomputer SQUID under the support of the D3 Center, Osaka University.

## AUTHOR DECLARATIONS
### Conflict of Interest

The authors have no conflicts to disclose.

### Author Contributions

**D. Pan:** Investigation (equal); Writing – original draft (equal); Writing – review & editing (equal). **M. Murakami:** Supervision (lead).

## DATA AVAILABILITY

The datasets used and/or analyzed during the current study are available from the corresponding author on reasonable request.